\documentclass[a4paper]{jpconf}

\usepackage{graphicx}
\usepackage{amsmath}
\begin{document}
%
%
\title{The remnant CP transformation and its implications }
%
%
\author{Felix Gonzalez-Canales}
%
%
\address{Departamento de F\'{\i}sica, 
 Centro de Investigaci\'on y de Estudios Avanzados del 
 Instituto Polit\'ecnico Nacional, 
 Apartado Postal 14-740, CDMX 07000, M\'exico.}
%
%
\ead{fgonzalez@fis.cinvestav.mx}
%
%
\begin{abstract}
 In the context of remnant CP transformations, I briefly discuss a generalized $\mu-\tau$ reflection symmetry,
 where the ``Majorana'' phases have CP conserving values, which are directly related with the CP 
 parities of neutrino states.  
 Also, one finds that the ``Dirac-like'' CP violation phase is correlated with the atmospheric 
 mixing angle, giving important phenomenological implications for current and future long baseline oscillation 
 neutrino experiments.     
\end{abstract}
%
%
\section{Introduction}
%
%
Nowadays it is well known that the flavour oscillations phenomenon is possible only if neutrinos have mass. 
This result contradicts the prediction of the Standard Model and is one of the strongest evidence that there 
must be a more complete theory. 
One of the main aims of theoretical physicists, is to identify a theory that responds to the question of 
existence of very small neutrino masses and, at the same time, explain the measured values of 
the parameters that govern its oscillations. 
At present, we are in the precision era for the determination of lepton mixing angles. However, it is not the 
same situation for the CP violation phase factors.

A non-zero reactor mixing angle, $\theta_{13}$, implies the existence of CP violation in lepton sector or 
vice versa.
Hence, one of the main goals of long baseline neutrino oscillation experiments  
(T2K~\cite{Abe:2015awa}, NO$\nu$A~\cite{Bian:2015opa} and DUNE~\cite{Acciarri:2015uup})
is the determination of the ``Dirac-like'' CP violation phase.
Also, $\theta_{13} \neq 0$ reduces the viability of several symmetry groups to be considered 
as  flavour symmetries in Standard Model extensions.

In agreement with the last results of global fits of neutrino oscillation data~\cite{Forero:2014bxa}, 
the elements of second and third row of lepton mixing matrix, satisfy the following approximate 
relation $\left| U_{\mu i} \right| \simeq \left| U_{\tau i} \right|$ with $i=1,2,3$. 
Then, so called $\mu-\tau$ symmetry is obtained if $\left| U_{\mu i} \right| = \left| U_{\tau i} \right|$ 
exactly holds.
We obtain this equality if and only if one of following two sets of conditions is satisfied: 
$\theta_{23} = \frac{\pi}{4}$ and $\theta_{13} = 0$, or 
$\theta_{23} = \frac{\pi}{4}$ and $\delta_{\mathrm{CP}} = \pm \frac{\pi}{2}$.
From the first set of conditions, one can see that neutrino mass term is invariant under the transformations: 
$\nu_{e} \to \nu_{e}$, $\nu_{\mu} \to \nu_{\tau}$ and $\nu_{\tau} \to \nu_{\mu}$, 
which are called $\mu-\tau$~{\it permutation symmetry}~\cite{Xing:2015fdg}. 
However, this kind of transformations is disfavoured by neutrino oscillation data, 
since $\theta_{13}$ is non-zero~\cite{Forero:2014bxa}. 
The last set of conditions implies that neutrino mass term is invariant under the transformations: 
$\nu_{e} \to \nu_{e}^{c}$, $\nu_{\mu} \to \nu_{\tau}^{c}$ and $\nu_{\tau} \to \nu_{\mu}^{c}$, 
where $c$ denotes the charge conjugation of neutrino field. 
This kind of transformations are called $\mu-\tau$~{\it reflection symmetry}~\cite{Xing:2015fdg}. 
Also, the values $\theta_{23} = \frac{\pi}{4}$ and $\delta_{\mathrm{CP}} = \pm \frac{\pi}{2}$ are allowed at 
$3\sigma$ level, while $\theta_{23} = \frac{\pi}{4}$ and $\delta_{\mathrm{CP}} = - \frac{\pi}{2}$ are allowed 
at $1\sigma$ or $2\sigma$ level~\cite{Forero:2014bxa}.

In this work, the residual CP transformations are used for redefining the flavour lepton mixing matrix. 
Thus, in the theoretical framework of a generalization of $\mu-\tau$ reflection symmetry, 
the possible values of CP phases are constrained, and its phenomenological implication for current and future 
long baseline neutrino oscillation experiments is shown.
%
%
\section{The hidden symmetries in the fermion mass matrices}
%
%
In order to construct the mass term of known neutrinos at low energy, here it is considered that neutrinos are 
Majorana particles. 
The leptonic flavour mixing implied by the neutrino oscillation phenomena can be described with help of the
Lagrangian 
\begin{equation}\label{Eq:Lagran:1}
 {\cal L} = 
  - \frac{ g }{ \sqrt{2} } \, \bar{\ell}_{L} \gamma^{\mu} \nu_{L} \, W_{\mu} 
  - \frac{1}{2} \nu_{L}^{\top} {\bf C}^{-1} {\bf M}_{\nu} \, \nu_{L} 
  - \bar{\ell}_{R} \, {\bf M}_{\ell} \, \ell_{L}
  + \mathrm{h. \, c.} \, ,
\end{equation}
where the charged lepton and Majorana neutrino mass terms are included. 
In the above expression the $L$ and $R$ subscripts denote the left- and right-handed field respectively, 
while 
$\ell_{L} = \left( e_{L}, \mu_{L}, \tau_{L} \right)^{\top}$ and 
$\ell_{R} = \left( e_{R}, \mu_{R}, \tau_{R} \right)^{\top}$ are the charged lepton fields, 
and finally 
$\nu_{L}  = \left( \nu_{e L}, \nu_{\mu L}, \nu_{\tau L} \right)^{\top}$ is the left-handed neutrino field. 
Note that the mass matrix ${\bf M}_{\nu}$ must be symmetric, since neutrinos are Majorana particles, 
while ${\bf M}_{\ell}$ does not have any special feature so it must be represented through a $3\times 3$ 
complex matrix. 
Respectively, the neutrino and charged lepton mass matrices can be brought to its diagonal form  
through the following unitary transformations
\begin{equation}\label{Eq:Mat:Diag}
 {\bf M}_{\nu}  = 
  {\bf U}_{\nu}^{*} \, {\bf M}_{\nu}^{\mathrm{diag}} \, {\bf U}_{\nu}^{ \dagger } \\
 \quad \textrm{and} \quad
 {\bf M}_{\ell} = 
  {\bf V}_{\ell} \, {\bf M}_{\ell}^{\mathrm{diag}} \, {\bf U}_{\ell}^{\dagger} \, ,
\end{equation}
where 
${\bf M}_{\nu}^{\mathrm{diag}} = \textrm{diag} \left( m_{\nu_{1}}, m_{\nu_{2}}, m_{\nu_{3}} \right)$
and 
${\bf M}_{\ell}^{\mathrm{diag}} = \textrm{diag} \left( m_{e}, m_{\mu}, m_{\tau} \right)$. 
The neutrino masses $m_{\nu_{i}}$ are real and non-negative. 
In agreement with the singular value decomposition theorem, ${\bf U}_{\ell}$ and ${\bf V}_{\ell}$ are the 
unitary matrices through which the Hermitian matrices ${\bf M}_{\ell} {\bf M}_{\ell}^{\dagger}$ and 
${\bf M}_{\ell}^{\dagger} {\bf M}_{\ell}$ can be brought to its diagonal form, respectively. 
The PMNS lepton mixing matrix  has the shape 
${\bf U}_{\mathrm{PMNS}} = {\bf U}_{\ell}^{\dagger} {\bf U}_{\nu}$~\cite{Hochmuth:2007wq}.
%
%
Now, since one is working in a independent model framework and without loss of generality, the charged lepton mass matrix will be considered with a diagonal shape. 
In the charged lepton diagonal basis the ${\bf U}_{\ell}$ unitary matrix is reduced to unit matrix, 
consequently the PMNS matrix takes the form ${\bf U}_{\mathrm{PMNS}} = {\bf U}_{\nu}$, 
which means that all information about the leptonic flavour mixing comes from the neutrino oscillations.
In order to find the flavour symmetry  hidden in the neutrino mass matrix is necessary to apply the following CP transformation of the left-handed neutrino fields to the lepton mass terms~\cite{Chen:2014wxa}
\begin{equation}
 \begin{array}{l}
  \nu_{L} \left( x \right) \rightarrow 
   i {\bf X}_{\nu} \, \gamma^{0} \, {\bf C} \, \bar{\nu}_{L}^{\top} \left( x_{p} \right)  ,
 \end{array}
\end{equation}
where ${\bf X}_{\nu}$ is a unitary matrix acting on family space, 
${\bf C}$ is the charge conjugation matrix, 
and $x_{p} = \left(t, -x \right)$. 
If the neutrino mass matrix satisfies the relation
\begin{equation}\label{Eq:Cond:Mat}
 {\bf M}_{\nu}^{*} = 
  {\bf X}_{\nu}^{\top} {\bf M}_{\nu} {\bf X}_{\nu} 
 \quad \Longrightarrow \quad 
 {\bf M}_{\nu} = 
  {\bf X}_{\nu}^{\dagger} {\bf X}_{\nu}^{\top} {\bf M}_{\nu} {\bf X}_{\nu} {\bf X}_{\nu}^{*} , 
\end{equation}
the corresponding mass term in the Lagrangian, eq.~(\ref{Eq:Lagran:1}), is invariant. 
The second expression in the last equation means that one flavour symmetry transformation is equivalent to 
successive application of two CP-transformations~\cite{Chen:2014wxa}.
From eqs.~(\ref{Eq:Mat:Diag}) and~(\ref{Eq:Cond:Mat}) we have
\begin{equation}\label{Eq:Sign:1}
 \begin{array}{l}
  {\bf U}_{\nu}^{\dagger} \, {\bf X}_{\nu} \, {\bf U}_{\nu}^{*} = 
   e^{i n_{1} \pi} \textrm{diag} \left( 1, e^{i n_{21} \pi},  e^{i n_{31} \pi} \right) ,
 \end{array}
\end{equation}
where $n_{21} = n_{2} - n_{1}$ and $n_{31} = n_{3} - n_{1}$ with $n_{i}$ elements of natural numbers. 
The 
phase factor $e^{i n_{1} \pi}$ can be absorbed by neutrino fields, so there exist four possible combinations 
of signs, which can be expressed as
\begin{equation}\label{Eq:X:U}
 \left( {\bf X}_{\nu} \right)_{i} = {\bf U}_{\nu} \, {\bf d}_{i} \, {\bf U}_{\nu}^{\top} ,
\end{equation}
where 
${\bf d}_{1} = \textrm{diag} \left( 1 , -1, -1 \right)$, 
${\bf d}_{2} = \textrm{diag} \left( 1 , -1,  1 \right)$,
${\bf d}_{3} = \textrm{diag} \left( 1 ,  1, -1 \right)$, and 
${\bf d}_{4} = \textrm{diag} \left( 1 ,  1,  1 \right)$.
%
%
From eq.~(\ref{Eq:Sign:1}) or~(\ref{Eq:X:U}) it is easy to conclude that the residual flavour 
symmetry of Majorana neutrinos mass matrix is the Klein group, namely $Z_2 \otimes Z_2$.
A direct consequence of eq.~(\ref{Eq:X:U}) is that ${\bf X}_{\nu} = {\bf X}_{\nu}^{\top}$. 
Therefore, the ${\bf X}_{\nu}$ is a unitary symmetric matrix. 
Applying the Takagi factorization to 
${\bf X}_{\nu}$ matrix, the eq.~(\ref{Eq:X:U}) can be written as
${\bf O}_{\nu} = {\bf O}_{\nu}^{*}$ or ${\bf O}_{\nu}^{\top} = {\bf O}_{\nu}^{\dagger}$, where  
{\small 
${\bf O}_{\nu} = {\bf \Sigma}^{\top} {\bf U}_{\nu} \textrm{diag} 
\left( e^{ -i \frac{ n_{1} }{2} \pi}, e^{ -i \frac{ n_{2} }{2} \pi}, e^{ -i \frac{ n_{3} }{2} \pi} 
\right)$}~\cite{Chen:2015siy,Chen:2016ica}.
Also,  ${\bf O}_{\nu} \, {\bf O}_{\nu}^{\dagger} = {\bf O}_{\nu} \, {\bf O}_{\nu}^{\top} = {\bf I}$, 
where ${\bf I}$ is the unit matrix. So it is very easy conclude that ${\bf O}_{\nu}$ matrix is 
a real orthogonal matrix. 
In this context the lepton flavour mixing matrix takes the shape~\cite{Chen:2015siy,Chen:2016ica}
\begin{equation}\label{Eq:PMNS:New}
 {\bf U}_{\mathrm{PMNS}} = {\bf U}_{\nu} =
  {\bf \Sigma}_{\nu} \, {\bf O}_{\nu} \, {\bf Q}_{\nu} ,
\end{equation} 
where 
\begin{equation}
 {\bf O}_{\nu} = 
 \left( \begin{array}{ccc}
   c_2 c_3 & c_2 s_3 & s_2 \\
  -c_3 s_1 s_2-c_1 s_3 & c_1 c_3-s_1 s_2 s_3 & c_2 s_1 \\
   s_1 s_3-c_1 c_3 s_2 & -c_3 s_1-c_1 s_2 s_3 & c_1 c_2 \\
 \end{array} \right)
\end{equation}
and 
${\bf Q}_{\nu} = \textrm{diag} 
  \left( e^{ -i \frac{ n_{1} }{2} \pi}, e^{ -i \frac{ n_{2} }{2} \pi}, e^{ -i \frac{ n_{3} }{2} \pi} \right)$.
Here, $c_{i} = \cos \theta_{i}$ and $s_{i} = \sin \theta_{i}$.
%
\section{Generalized $\mu-\tau$ symmetry}
%
Here a generalized $\mu-\tau$ reflection symmetry is proposed, defined as~\cite{Chen:2015siy,Chen:2016ica}:
\begin{equation}
 {\bf X}_{\nu} = 
 \left( \begin{array}{ccc}
  e^{i \alpha} & 0 & 0 \\
  0 & e^{i \beta} \cos \Theta & i e^{i \frac{ \left( \beta + \gamma \right) }{ 2 } } \sin \Theta \\
  0 & i e^{i \frac{ \left( \beta + \gamma \right) }{ 2 } } \sin \Theta & e^{i \gamma} \cos \Theta
 \end{array}  \right)
 \xrightarrow[\Theta = \pm \frac{\pi}{2} ]{\alpha = 0, \; \beta = 0, \; \gamma = 0}  
 \left( \begin{array}{ccc}
  1 & 0 & 0 \\
  0 & 0 & i \kappa \\
  0 & i \kappa & 0
 \end{array}  \right) ,
\end{equation} 
where $\kappa = \pm 1$, the phase factors $\alpha$, $\beta$, $\gamma$, and the angle $\Theta$ are real 
parameters. 
In the particular case of having $\alpha = 0$, $\beta = 0$, $\gamma = 0$, and 
$\Theta = \pm \frac{\pi}{2}$, 
the ${\bf X}_{\nu}$ matrix is reduced to standard $\mu-\tau$ reflection symmetry~\cite{Xing:2015fdg}. 
The Takagi factorization of ${\bf X}_{\nu}$ matrix has the shape 
\begin{equation}
 {\bf \Sigma}_{\nu} = 
 \left( \begin{array}{ccc}
  e^{i \frac{ \alpha }{2} } & 0 & 0 \\
  0 & e^{i \frac{ \beta }{2} } \cos \Theta & i e^{i \frac{ \beta }{2} } \sin \Theta \\
  0 & i e^{i \frac{ \gamma }{2} } \sin \Theta & e^{i \frac{ \gamma }{2} } \cos \Theta
 \end{array}  \right) .
\end{equation} 

If the ${\bf \Sigma}_{\nu}$ matrix has the form given in the above expression, 
the relation between flavour mixing angles and the entries of the lepton mixing matrix, 
eq.~(\ref{Eq:PMNS:New}), are: 
\begin{equation}
 \begin{array}{l}\vspace{2mm}
 \sin^{2} \theta_{12} = 
  \frac{ \left| U_{e2} \right|^{2} 
  }{  
   1 - \left| U_{e3} \right|^{2} 
  } = \sin^{2} \theta_{3},  
 \quad
 \sin^{2} \theta_{13} = \left| U_{e3} \right|^{2} = \sin^{2} \theta_{2}, \\
 \sin^{2} \theta_{23} =  
  \frac{ \left| U_{\mu 3} \right|^{2} 
  }{  
   1 - \left| U_{e3} \right|^{2} 
  }
  = \frac{1}{2} \left( 1 - \cos \Theta \cos 2\theta_{1} \right).
 \end{array}
\end{equation}

The phase factors associated to CP violation are determined through the Jarlskog invariant 
{\small${\cal J}_{\mathrm{CP}} = {\cal I}m \left \{ U_{e1}^{*} U_{\mu 3}^{*} U_{e3} U_{\mu 1} \right \}$} and
the invariants 
{\small$\textrm{I}_{1} = {\cal I}m \left \{ U_{e2}^{2} U_{e1}^{*2} \right \}$} and 
{\small$\textrm{I}_{2} = {\cal I}m \left \{ U_{e2}^{2} U_{e1}^{*2} \right \}$}. 
The explicit form of these phases are~\cite{Chen:2015siy}:
\begin{equation}
 \begin{array}{l}
  \sin \delta_{\mathrm{CP}} = 
   \frac{ 
    \sin \Theta \, \textrm{sign} \left[ \sin \theta_{2} \sin 2 \theta_{3} \right] 
   }{  
    \sqrt{ 1 - \cos^{2} \Theta \cos^{2} 2\theta_{1} } 
   } , \qquad
  \phi_{12} = \frac{ n_{21} }{2} \pi 
  \quad \textrm{and} \quad
  \phi_{13} = \frac{ n_{31} }{2} \pi .
 \end{array}
\end{equation}
Here, $\delta_{\mathrm{CP}}$ is the Dirac-like phase which is involved in the neutrino oscillation. 
Therefore, this phase can be measured in the long baseline neutrino oscillation experiments. 
On the other hand, $\phi_{12}$ and $\phi_{13}$ are the Majorana phases which are involved in the 
neutrinoless double beta decay. 
The PMNS mixing matrix in the standard parametrization has six free parameters, three angles and three phase 
factors, while in the context of generalized $\mu-\tau$ reflection symmetry this matrix has just four free 
parameters, which are four angles. 
In our theoretical framework, we have no authentic prediction for the lepton mixing angles and Dirac-like CP 
violation phase. 
However, we have an important prediction for the Majorana phases. The $\phi_{12}$ and $\phi_{13}$ 
phases are CP conserving and have a direct relation with the CP parities of the neutrino 
states~\cite{Chen:2015siy}. 
%
%
\section{Neutrino oscillation in matter}
%
%
A viable way of observing CP violation in the leptonic sector is through the differences in the oscillation 
probabilities that involve neutrinos and antineutrinos. 
This difference in the vacuum has the form~\cite{Nunokawa:2007qh}: 
$\Delta P_{\alpha \beta} \equiv 
 P \left( \nu_{\alpha} \to \nu_{ \beta } \right) - P \left( \bar{\nu}_{\alpha} \to \bar{\nu}_{ \beta } \right) 
 = - 16 \, J_{\alpha \beta} \, \sin \Delta_{ 21 } \sin \Delta_{ 23 } \sin \Delta_{ 31 }$.
Here, $\Delta_{kj} = \Delta m_{kj}^{2} L /(4E)$ and $\Delta m_{kj}^{2} = m_{\nu_{k}}^{2} - m_{\nu_{j}}^{2}$, 
$L$ is the baseline, $E$ is the energy of neutrino beam, and ${\cal J}_{\alpha \beta}$ is the Jarlskog 
invariant in its leptonic version, whose definition is 
${\cal J}_{\alpha \beta} = 
 {\cal I}m \left( U_{\alpha 1} U_{\beta 2} U^{\ast}_{\alpha 2} U^{\ast}_{\beta 1}\right) = \pm J_{CP}$. 
The positive sign correspond to a cyclic permutation of the flavour indices  
$e$, $\mu$ and $\tau$, while the negative sign is for an anti-cyclic permutation.

The present long baseline experiments like T2K and NO$\nu$A as well as the proposed experiment DUNE, are 
interested in the measurements of the parameters that govern the $\nu_{\mu} \to \nu_{e}$ oscillation.
In the vacuum, the transition probabilities 
$P \left( \nu_{\mu} \to \nu_{e} \right)$ and $P \left( \bar{\nu}_{\mu} \to \bar{\nu}_{e} \right)$ have the 
form~\cite{Nunokawa:2007qh}
\begin{equation}\label{Eq:Vacuum:Pmu-e}
 \begin{array}{l}\vspace{2mm}
  P \left( \nu_{\mu} \to \nu_{e} \right) \simeq 
   P_{\mathrm{atm}} 
   + 2 \sqrt{P_{\mathrm{atm}}} \sqrt{P_{\mathrm{sol}}} \cos\left( \Delta_{32} + \delta_{\mathrm{CP}} \right) 
   + P_{ \mathrm{sol} }\,, \\
  P \left( \bar{\nu}_{\mu} \to \bar{\nu}_{e} \right) \simeq 
   P_{\mathrm{atm}} 
   + 2 \sqrt{P_{\mathrm{atm}}} \sqrt{P_{\mathrm{sol}}} \cos\left( \Delta_{32} - \delta_{\mathrm{CP}} \right) 
   + P_{ \mathrm{sol} }\,,
 \end{array}
\end{equation}
where
$\sqrt{ P_{ \mathrm{atm} } } = \sin \theta_{23} \sin 2 \theta_{13} \sin \Delta_{31}$ and 
$\sqrt{ P_{ \mathrm{sol} } } = \cos \theta_{23} \cos \theta_{13} \sin 2\theta_{12} \sin \Delta_{21}$.
Then, the asymmetry between transition probabilities 
$P \left( \nu_{\mu} \to \nu_{e} \right)$ and $P \left( \bar{\nu}_{\mu} \to \bar{\nu}_{e} \right)$ is given by
\begin{equation}\label{Eq:Asym:e_mu:1}
 A_{\mu e} = 
 \frac{ 
  P \left( \nu_{\mu} \to \nu_{e} \right) - P \left( \bar{\nu}_{\mu} \to \bar{\nu}_{e} \right) 
 }{ 
  P \left( \nu_{\mu} \to \nu_{e} \right) + P \left( \bar{\nu}_{\mu} \to \bar{\nu}_{e} \right)
 } =
 - \frac{ 
  2 \sqrt{P_{ \mathrm{atm} } } \sqrt{ P_{\mathrm{sol}} } \sin \Delta_{32} \sin\delta_{\mathrm{CP}} 
 }{ 
  P_{\mathrm{atm}} + 2 \sqrt{ P_{\mathrm{atm}} } \sqrt{ P_{\mathrm{sol}} } \cos \Delta_{32} 
  \cos \delta_{\mathrm{CP}} + P_{\mathrm{sol}} 
 } \,.
\end{equation}
However, to make a realistic description of $\nu_{\mu} \to \nu_{e}$ oscillation in the long baseline 
experiments, it is necessary to include the matter effects associated with neutrino propagation inside the 
Earth. 
Then, the expressions for $\sqrt{P_{\mathrm{atm}}}$ and $\sqrt{P_{\mathrm{sol}}}$ in matter 
are~\cite{Nunokawa:2007qh}:
\begin{equation}\label{Eq:Pmu-e}
 \sqrt{ P_{\mathrm{atm}} } = 
  \sin \theta_{23} \sin 2\theta_{13} \frac{ \sin \left( \Delta_{31} - aL \right) }{ \Delta_{31} - aL } \,
  \Delta_{31} \,,\quad 
 \sqrt{ P_{\mathrm{sol}} } = 
  \cos \theta_{23} \sin 2\theta_{12} \frac{ \sin(aL) }{ aL } \, \Delta_{21}\,.
\end{equation}
Here, $a=G_{F}N_{e}/\sqrt{2}$, where $G_{F}$ is the Fermi constant and $N_{e}$ is the density of electrons. 
For the earth crust the parameter $a$ is approximately  
$a \approx (3500\mathrm{km})^{-1}$~\cite{Chen:2016ica}. 
The neutrino transition probability, $P \left( \nu_{\mu} \to \nu_{e} \right)$, is obtained by replacing the 
eq.~(\ref{Eq:Pmu-e}) into first expression of eq.~(\ref{Eq:Vacuum:Pmu-e}). 
However, for transition probability $P \left( \bar{\nu}_{\mu} \to \bar{\nu}_{e} \right)$ 
we can not only make the previous substitution, because the involved particles are the antineutrinos and 
we need make the change $a \to -a$. 
Therefore, we make the redefinitions $\sqrt{P_{\mathrm{atm}}} \to \sqrt{ {\cal P}_{\mathrm{atm}}} $ and 
$\sqrt{P_{\mathrm{sol}}} \to \sqrt{ {\cal P}_{\mathrm{sol}}}$, where 
\begin{equation}\label{Eq:bar_Pmu-e}
 \sqrt{ {\cal P}_{\mathrm{atm}} } = 
  \sin \theta_{23} \sin 2\theta_{13} \frac{ \sin \left( \Delta_{31} + aL \right) }{ \Delta_{31} + aL } \,
  \Delta_{31} \,,\quad 
 \sqrt{ {\cal P}_{\mathrm{sol}} } = 
  \cos \theta_{23} \sin 2\theta_{12} \frac{ \sin(aL) }{ aL } \, \Delta_{21}\,.
\end{equation}
The antineutrino oscillation probability $P(\bar{\nu}_{\mu}\rightarrow\bar{\nu}_{e})$ is 
obtained replacing the eq.~(\ref{Eq:bar_Pmu-e}) in second expression of eq.~(\ref{Eq:Vacuum:Pmu-e}).
Finally, the asymmetry between transition probabilities  
$P \left( \nu_{\mu} \to \nu_{e} \right)$ and $P \left( \bar{\nu}_{\mu} \to \bar{\nu}_{e} \right)$ 
including matter effects takes the form
\begin{equation}\label{Eq:Asym:e_mu:2}
 \begin{array}{l}
 {\cal A}_{\mu e} =
 \frac{ 
  \left( P_{\mathrm{atm}} - {\cal P}_{\mathrm{atm}} \right) + 2 \sqrt{ P_{\mathrm{sol}} } 
  \left[ \left( \sqrt{ P_{\mathrm{atm}} } - \sqrt{ {\cal P}_{\mathrm{atm}} } \right) \cos \Delta_{32} 
  \cos \delta_{\mathrm{CP}} 
  - \left( \sqrt{ P_{\mathrm{atm}} } + \sqrt{ {\cal P}_{\mathrm{atm}} } \right) \sin \Delta_{32} 
  \sin \delta_{\mathrm{CP}} \right]
 }{ 
  2 P_{\mathrm{sol}} + \left( P_{\mathrm{atm}} + {\cal P}_{\mathrm{atm}} \right) + 2 \sqrt{ P_{\mathrm{sol}} }
  \left[ \left( \sqrt{ P_{\mathrm{atm}} } + \sqrt{ {\cal P}_{\mathrm{atm}} } \right) \cos \Delta_{32} 
  \cos \delta_{\mathrm{CP}} 
  - \left( \sqrt{ P_{\mathrm{atm}} } - \sqrt{ {\cal P}_{\mathrm{atm}} } \right) \sin \Delta_{32} 
  \sin \delta_{\mathrm{CP}} \right] 
 }.
 \end{array}
\end{equation}
\begin{figure}[!htbp]
 \begin{center}
  \begin{tabular}{cc}
   \includegraphics[scale=.6]{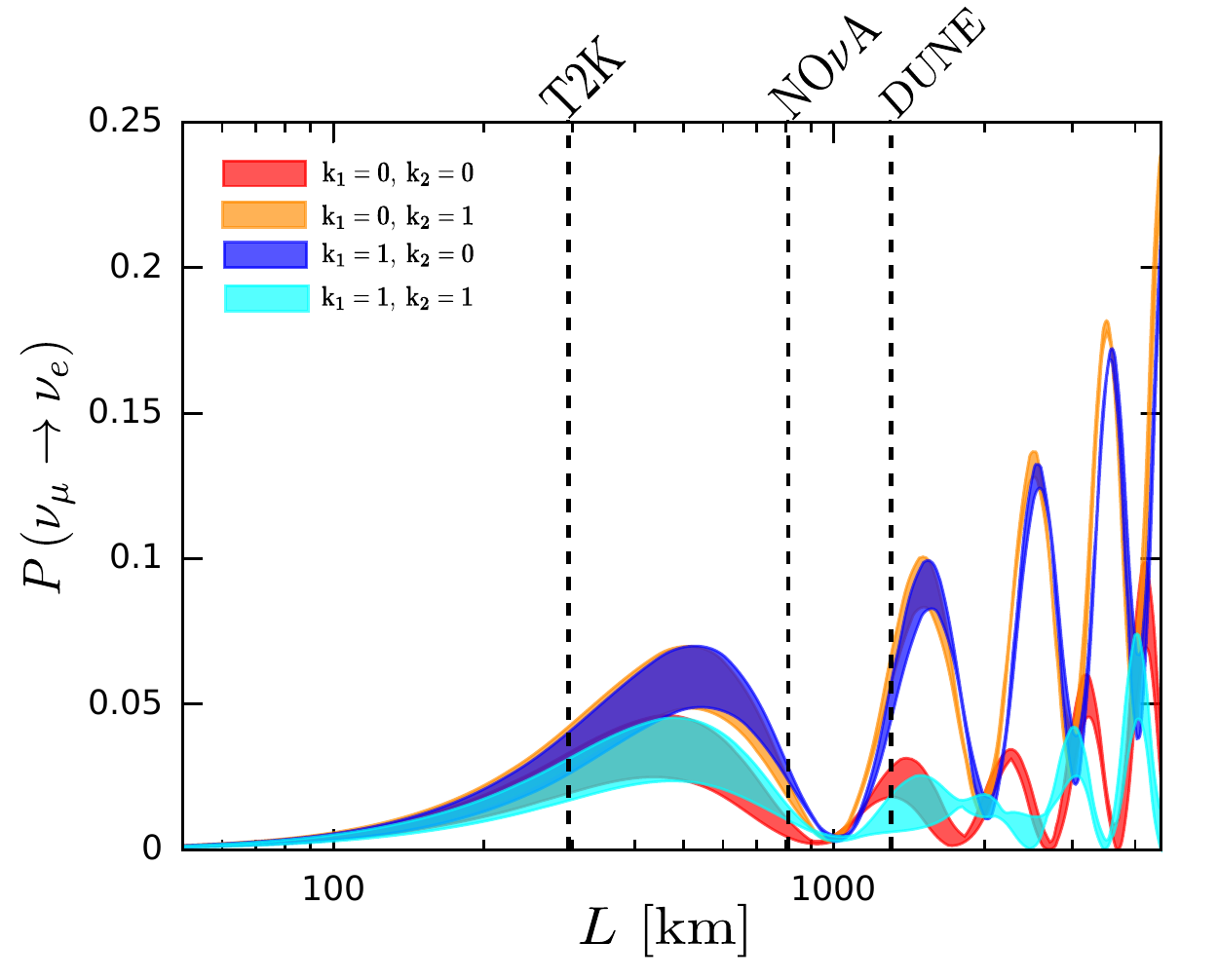} & 
   \includegraphics[scale=.6]{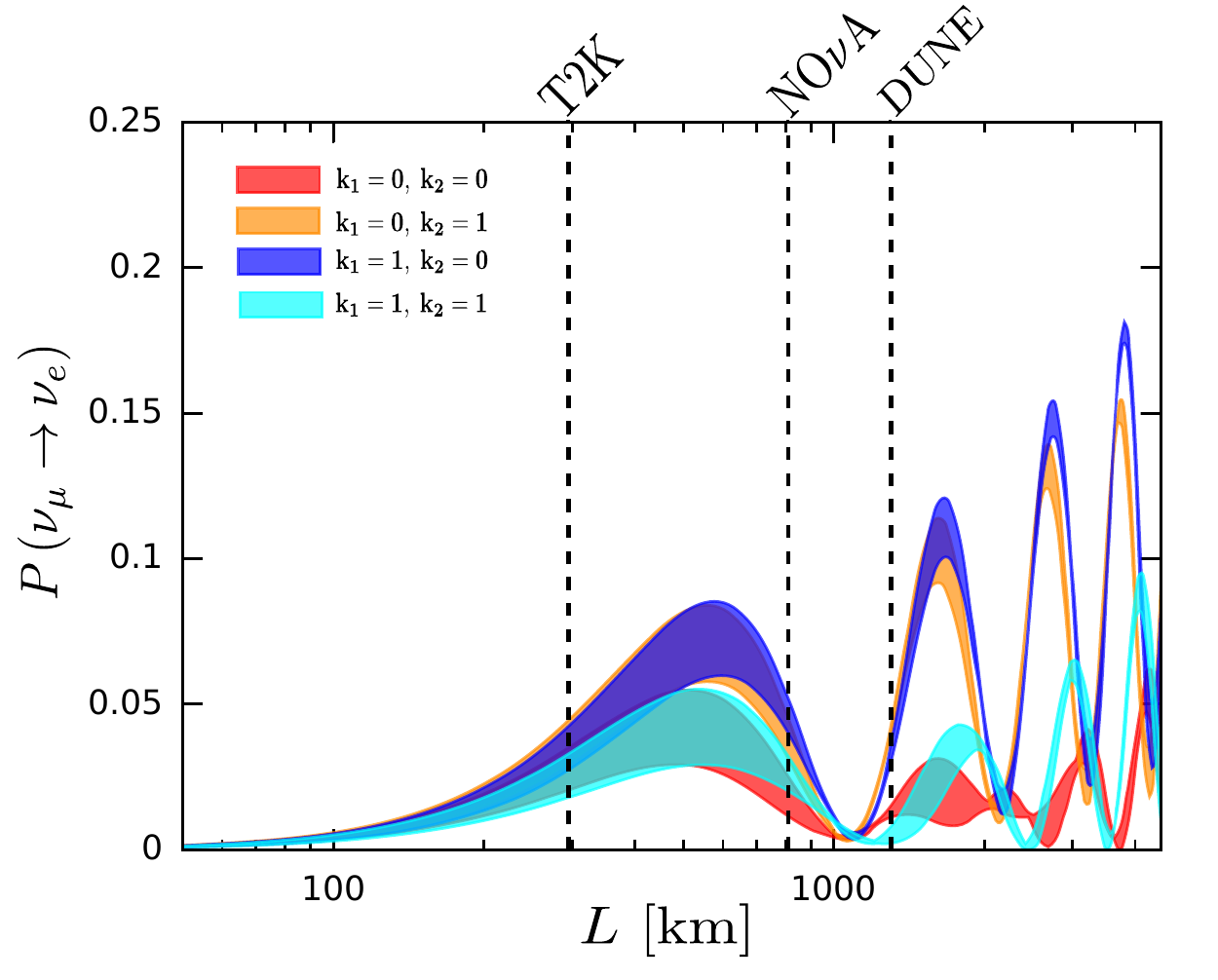} \\ 
   \includegraphics[scale=.6]{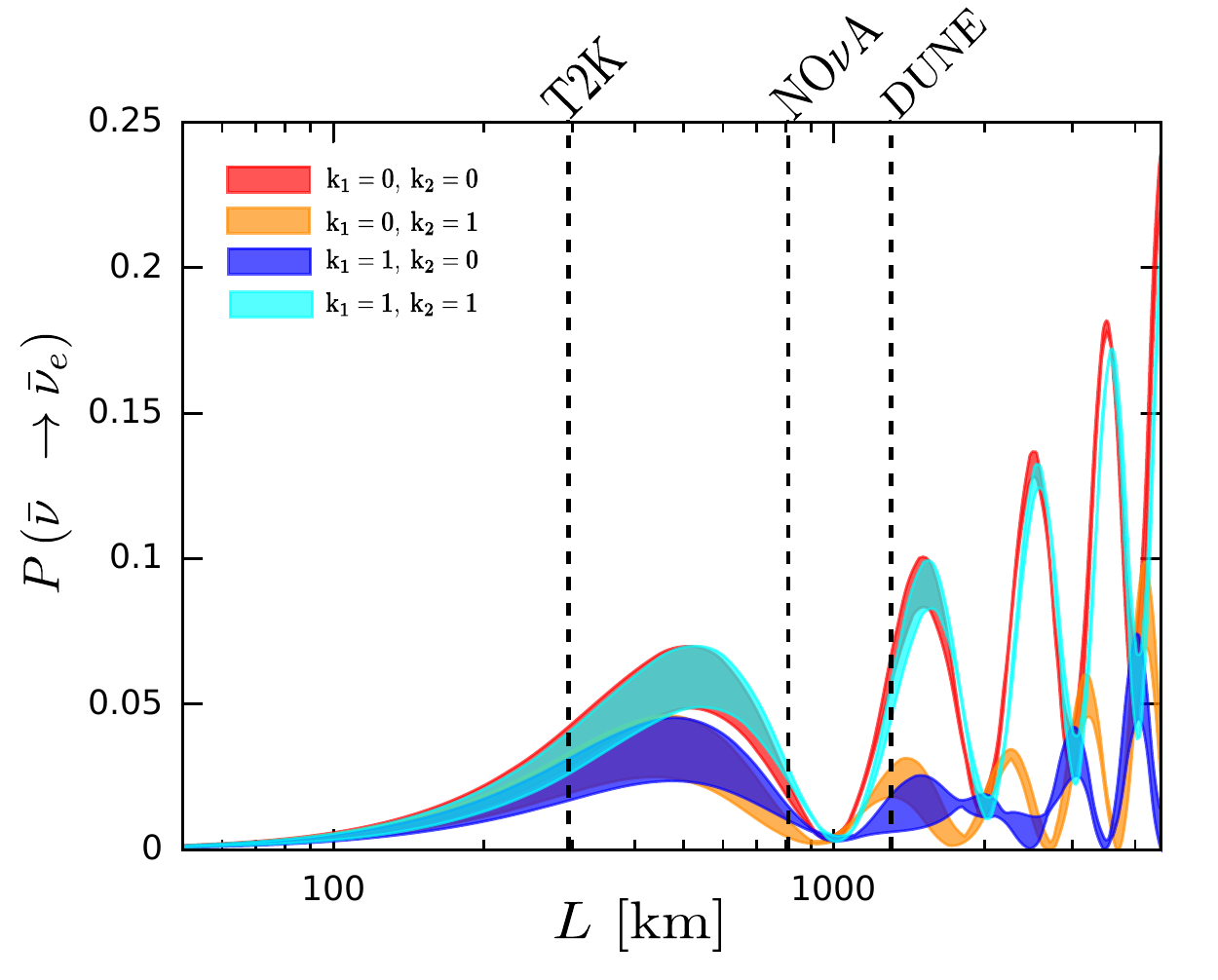} & 
   \includegraphics[scale=.6]{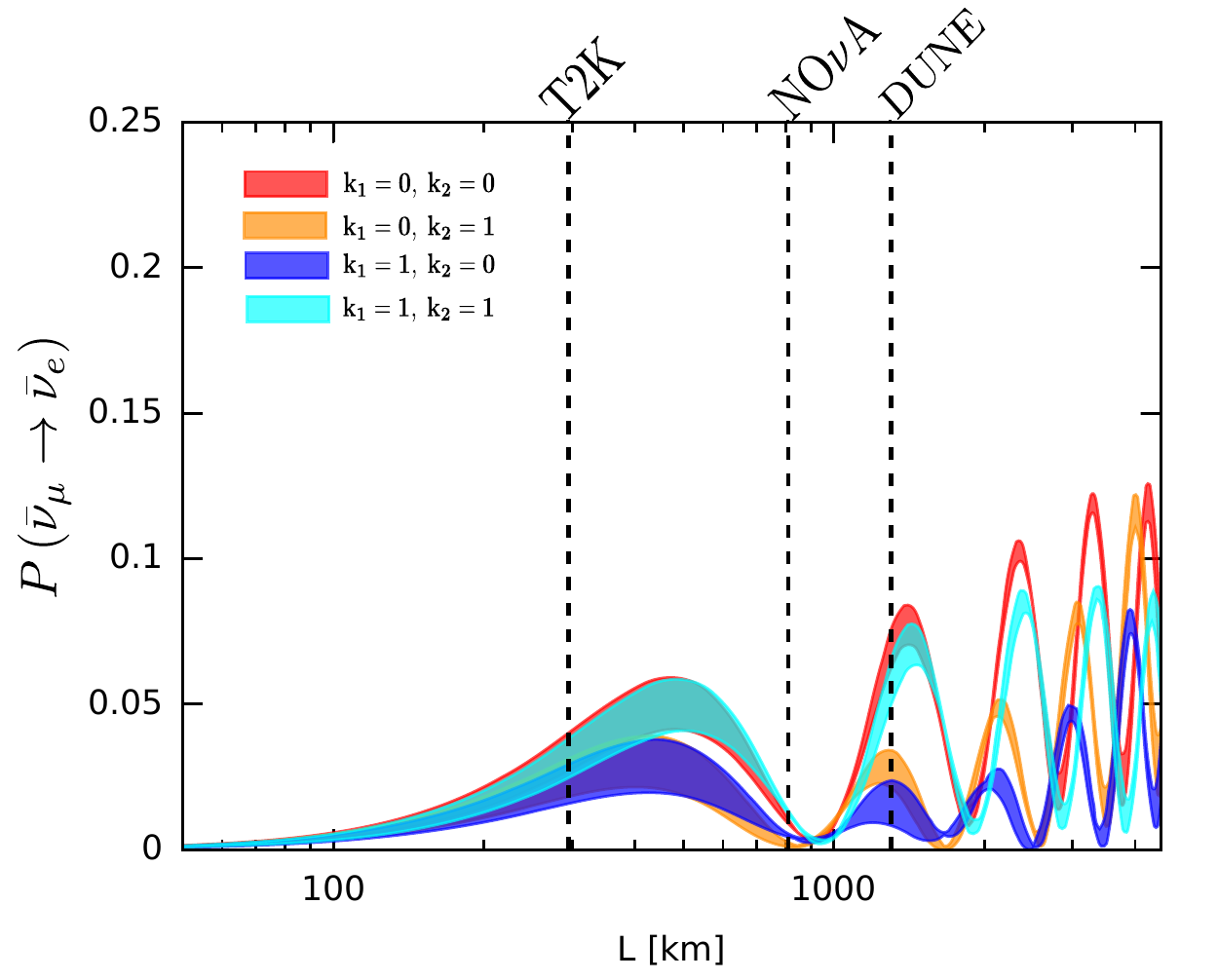} \\ 
   \includegraphics[scale=.6]{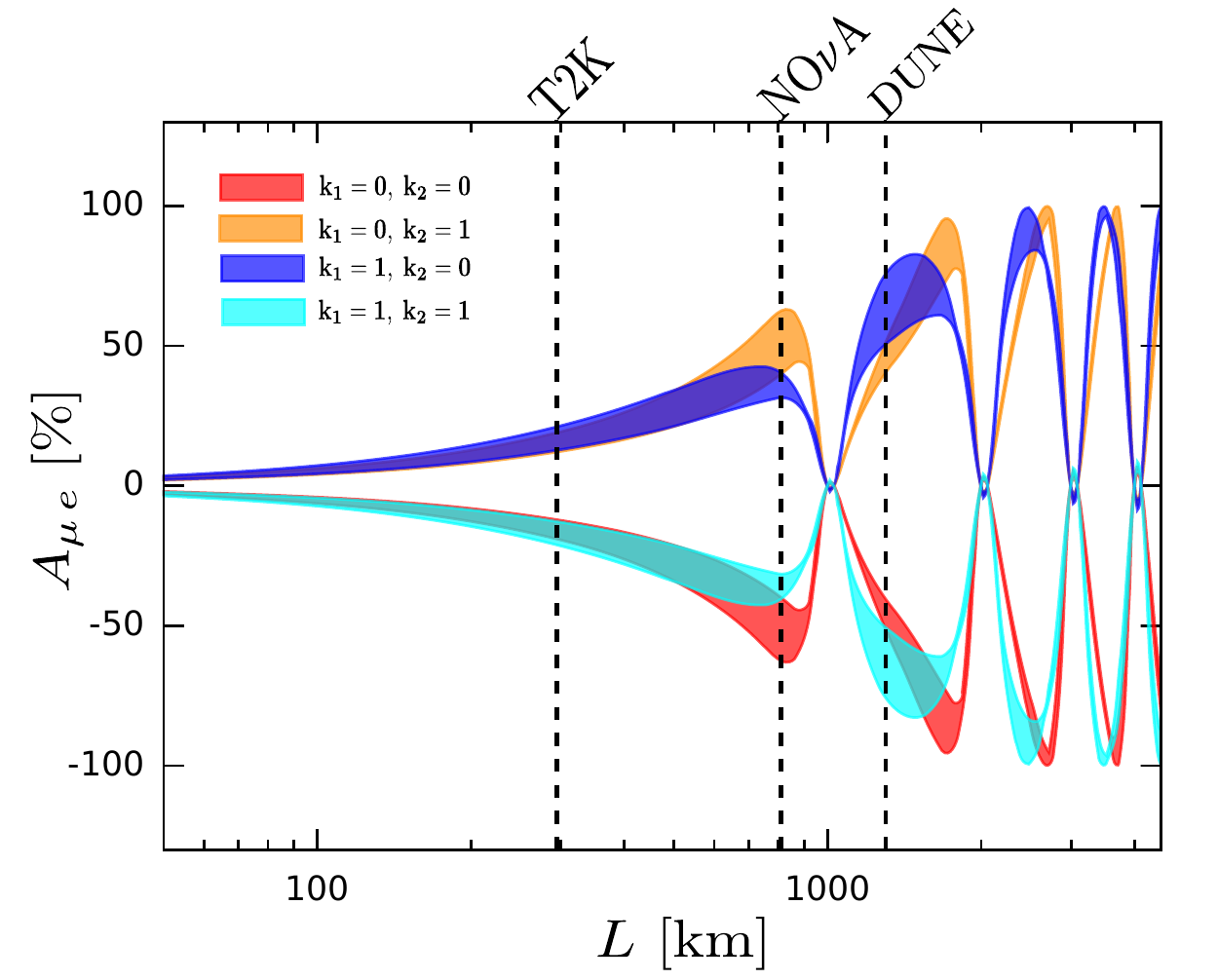} &
   \includegraphics[scale=.6]{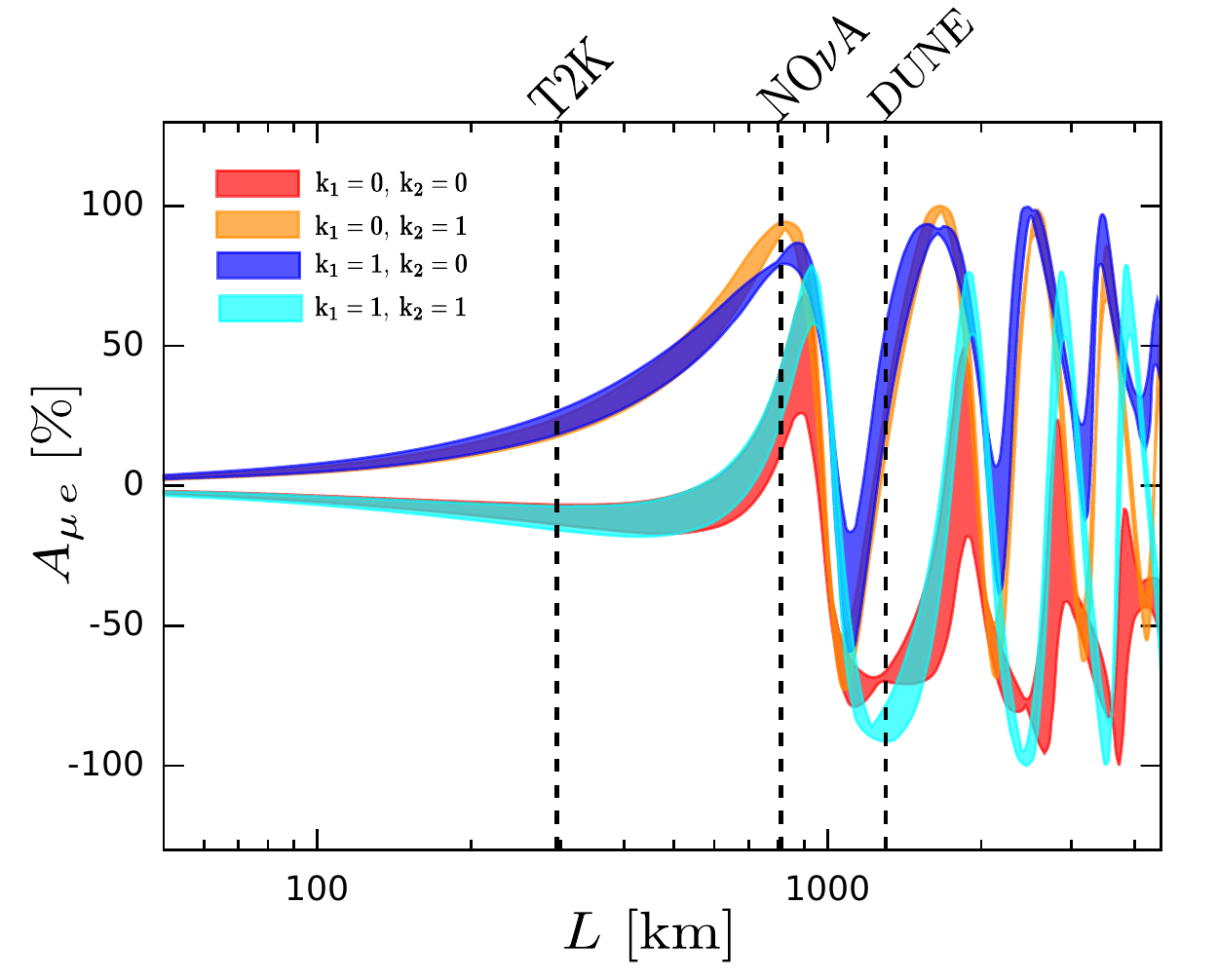}
 \end{tabular}
 \caption{\small
  The transition probabilities $\nu_{\mu} \to \nu_{e}$ and $\bar{\nu}_{\mu} \to \bar{\nu}_{e}$, 
  as well as the $A_{\mu e}$ and ${\cal A}_{\mu e}$ asymmetries for a energy of neutrino beam $E = 1$~GeV.
  Respectively, the left and right panels correspond to neutrino oscillation in vacuum and matter.
  For a normal hierarchy in the neutrino mass spectrum, the atmospheric mixing angle is taken within 
  3$\sigma$ range $0.393 \leq sin^{2} \theta_{23} \leq 0.643$~\cite{Forero:2014bxa}. 
  The other neutrino oscillation parameters are taken at their best fit point values: 
  $\Delta m_{21}^{2}    = 7.60 \times 10^{-5}~\textrm{eV}^{2}$,
  $|\Delta m_{31}^{2}|  = 2.48 \times 10^{-3}~\textrm{eV}^{2}$,
  $\sin^{2} \theta_{12} = 0.323$ and $\sin^{2} \theta_{13} = 0.0226$, 
  while $\Theta$ parameter is $3\pi/8$.
  }\label{Figura}
 \end{center}
\end{figure}

In the context of generalized $\mu-\tau$ reflection symmetry, the Dirac-like phase, atmospheric mixing 
angle and $\Theta$ parameter that characterizes the ${\bf X}_{\nu}$ CP transformation matrix, are 
related each other through the expression
\begin{equation}\label{Eq:d_cp:mu-tau}
 \sin^{2} \delta_{\mathrm{CP}} \sin^{2} 2\theta_{23} = \sin^{2} \Theta
 \quad \Rightarrow \quad
 \delta_{\mathrm{CP}} = 
  k_{1} \, \pi + (-1)^{k_{2}} \arcsin \left( \frac{ \sin \Theta }{ \sin 2\theta_{23} } \right),
\end{equation}
where $k_{1,2} = 0,1$.
This correlation allow us to obtain the value range of the Dirac-like CP violation phase. 
In the particular case when the $\Theta$ angle is $\pm \pi/4$, the other two parameters, $\theta_{23}$ and 
$\delta_{\mathrm CP}$, involved in the above expression are maximal. 
In agreement with the results of global fits of neutrino oscillation data~\cite{Forero:2014bxa}, a maximal 
atmospheric angle is disfavoured. Consequently, the called standard $\mu-\tau$ reflection symmetry is 
disfavoured. So, a non-maximal atmospheric mixing angle makes that our generalized $\mu-\tau$ reflection 
symmetry scenario be a good alternative for the CP violation study. 
The explicit form of $\nu_{\mu} \to \nu_{e}$ and $\bar{\nu}_{\mu} \to \bar{\nu}_{e}$ transition probabilities, 
as well as the asymmetry between these transition probabilities, are obtained by substituting the second 
expression of eq.~(\ref{Eq:d_cp:mu-tau}) in eqs.~(\ref{Eq:Vacuum:Pmu-e}),~(\ref{Eq:Asym:e_mu:1}) and~(\ref{Eq:Asym:e_mu:2}).
For instance, the $ P \left( \nu_{\mu} \to \nu_{e} \right)$ and 
$ P \left( \bar{\nu}_{\mu} \to \bar{\nu}_{e} \right)$ transition probabilities in matter have the following 
explicit shape  
{\small
\begin{equation}
 \begin{array}{l}\vspace{2mm}
  P \left( \nu_{\mu} \to \nu_{e} \right) \simeq 
   P_{\mathrm{atm}} 
   + 2 \sqrt{P_{\mathrm{atm}}} \sqrt{P_{\mathrm{sol}}} \cos \left( k_{1} \pi \right)
   \cos \left( 
    \Delta_{32} + ( -1 )^{k_{2}} \arcsin \left( \frac{ \sin \Theta }{ \sin 2\theta_{23} } \right) \right) 
   + P_{ \mathrm{sol} }\, , \\
  P \left( \bar{\nu}_{\mu} \to \bar{\nu}_{e} \right) \simeq 
   {\cal P}_{\mathrm{atm}} 
   + 2 \sqrt{ {\cal P}_{\mathrm{atm}}} \sqrt{P_{\mathrm{sol}}} \cos \left( k_{1} \pi \right)
   \cos \left( \Delta_{32} + (-1)^{k_{2} + 1 } \arcsin \left( \frac{ \sin \Theta }{ \sin 2\theta_{23} } \right) \right) 
   + P_{ \mathrm{sol} }\, .
 \end{array} 
\end{equation} } 
In the Fig.~\ref{Figura} is shown the behaviour of $\nu_{\mu} \to \nu_{e}$ and 
$\bar{\nu}_{\mu} \to \bar{\nu}_{e}$ transition probabilities, as well as $A_{\mu e}$ and ${\cal A}_{\mu e}$ 
asymmetries.

%
\section{Conclusions}
%
We proposed a generalized $\mu-\tau$ reflection symmetry where the PMNS mixing matrix has four free parameters 
only, in contrast with the six parameters in the standard parametrization. 
We obtained that the ``Majorana'' phases have CP conserving values which are directly related with the 
CP parities of neutrino states. 
On the other hand, the ``Dirac-like'' CP violation phase is correlated with the $\theta_{23}$ atmospheric 
mixing angle and $\Theta$ angle, which characterizes the CP transformation ${\bf X}_{\nu}$ matrix. 
This correlation has important implications for the long baseline oscillation neutrino experiments 
T2K, NO$\nu$A and DUNE. 
For a review of the phenomenological implications of our generalized $\mu-\tau$ reflection symmetry for the 
neutrinoless double beta decay and the leptogenesis see~\cite{Chen:2015siy,Chen:2016ica}.
%
\subsection*{Acknowledgments}
%
This talk was based on work in collaboration with P.~Chen, G.~J.~Ding and J.~W.~F.~Valle, and has been 
partially supported  by \textit{CONACYT-SNI (M\'exico)} and {\it CONACYT} under grant 236394. 
The author thanks the support provided by the Laboratorio Nacional de Superc\'omputo del Sureste de M\'exico 
through the grant number O-2016/039.

\end{document}